# Direct imaging of liquid-nanoparticle interface with atom probe tomography

Shi. Qiu[1], Changxi Zheng[2], Qi Zhou[3], Dashen Dong[4], Qianqian Shi[5], Vivek Garg[1,6], Shuo Zhang[1], Wenlong Cheng[5], Ross K.W. Marceau[7], Gang Sha[3*], Jing. Fu[1,8*]

[1]Department of Mechanical and Aerospace Engineering, Monash University, Clayton, VIC 3800, Australia

[2]Institute of Natural Sciences, Westlake Institute of Advanced Study, Westlake University, Hangzhou 310024, Zhejiang Province, China

[3]School of Materials Science and Engineering/Herbert Gleiter Institute of Nanoscience,

Nanjing University of Science and Technology, Xiaolingwei 200, Nanjing, Jiangsu, 210094 China

[4]Functional Materials and Microsystems Research Group and the Micro Nano Research Facility, RMIT University, Melbourne, VIC 3001, Australia

[5]Department of Chemical Engineering, Monash University, Clayton, VIC 3800, Australia

[6]IITB-Monash Research Academy, Indian Institute of Technology Bombay, Powai, Mumbai, 400076, India

[7]Deakin University, Institute for Frontier Materials, Geelong, VIC 3216, Australia

[8]ARC Centre of Excellence for Advanced Molecular Imaging, Monash University, Clayton, VIC 3800, Australia

*Corresponding authors. Email: gang.sha@njust.edu.cn or jing.fu@monash.edu

**Abstract**

Understanding the structure and chemical composition at the liquid-nanoparticle (NP) interface is crucial for a wide range of physical, chemical and biological processes. In this study, direct imaging of the liquid-NP interface by atom probe tomography (APT) is reported for the first time, which reveals the distributions and the interactions of key atoms and molecules in this critical domain. The APT specimen is prepared by controlled graphene encapsulation of the solution containing nanoparticles on a metal tip, with an end radius in the range of 50 nm to allow field ionization and evaporation. Using Au nanoparticles (AuNPs) in suspension as an example, analysis of the mass spectrum and three-dimensional (3D) chemical maps from APT provides a detailed image of the water-gold interface with near-atomic resolution. At the water-gold interface, the formation of an electrical double layer (EDL) rich in water ($H_2O$) molecules has been observed, which results from the charge from the binding between the trisodium-citrate layer and the AuNP. In the bulk water region, the density of reconstructed $H_2O$ has been shown to be consistent, reflecting a highly packed density of $H_2O$ molecules after graphene encapsulation. This study is the first demonstration of direct imaging of liquid-NP interface using APT with results providing an atom-by-atom 3D dissection of the liquid-NP interface.

# 1. Introduction

Investigating the liquid-nanoparticle (NP) interface plays a significant role in a wide range of chemical, physical and biological processes.[1-3] The interface has been studied using a spectrum of characterization techniques such as x-ray scattering, atomic force microscopy (AFM) and cryo-electron microscopy.[4-6] However, high-resolution chemical mapping of the interface particularly direct imaging of liquid molecules at the interface remains challenging due to a lack of characterization methods. Atom probe tomography (APT), as a unique technique providing three-dimensional (3D) chemical mapping of nanostructure with near-atomic resolution (~0.3 nm laterally and ~0.1 nm in depth),[7] holds great potentials to unravel this critical domain. The principle of APT is that, by applying high voltage to the needle-shaped specimen under cryogenic conditions and ultra-high vacuum, an electrostatic field can be created on the apex to achieve field ionization. Controlled high-voltage or laser pulses superimposed to the electrostatic field allows the layer-by-layer evaporation of surface atoms and molecules.[8] The evaporated ions are projected to a time-resolved and single-particle detector. The time-of-flight determines the ionic species, and the original positions of ions are reconstructed into 3D chemical maps based on the detector impact sequence and position using a reverse projection algorithm.[9] In order to achieve electrostatic field intensity (~10 V/nm) suitable for field ionization and evaporation, the APT specimen must have an end radius of less than 75 nm. The longstanding aspiration to image the liquid-NP interface by APT has been mainly impeded by sophisticated sample preparation. During conventional techniques for preparing APT specimens by site-specific focused ion beam (FIB) lift-out, dehydration of the samples is inevitable due to the vacuum environments involved.[10, 11] Notably, cryo-transfer accessories have been recently implemented to transport the frozen sample from cryo-FIB instrument to the atom probe instrument, and the overall approach has been demonstrated to be feasible to explore the liquid-solid interface.[12, 13] However, preparing NPs in solution with cryo-FIB lift-out followed by cryo-transfer to atom probe instrument still remains an extremely challenging and time-consuming task, and no such studies have been reported so far.

Recently, graphene has been employed to coat insulated specimens for voltage-pulsed APT imaging,[14] due to its outstanding conductivity and mechanical properties. Imaging liquid specimens has also been proved to be feasible with minimal damages to unravel the protein structure.[15] In this study, we further extend this technique and demonstrate that NPs in solution can be effectively encapsulated on pre-sharpened tips with graphene membranes. Gold nanoparticle (AuNP) has been selected considering that its inert characteristic can prevent the electrolysis or electrochemical reactions under a broad range of electrostatic potentials.[16] AuNP is also a significant anode material for rechargeable lithium (Li)-ion batteries, and the interface between solution and AuNP directly affects the capacity and cycle life.[17-19] Solution containing AuNPs can be effectively trapped on tungsten (W) tips with a final geometry suitable for APT by tuning the encapsulating speed and the number of graphene encapsulations (Figure 1a). Characterization by TEM (transmission electron microscope) and SEM (scanning electron microscope) is employed in parallel to confirm that the solution containing AuNPs can be trapped on pre-sharpened W tips. Upon being loaded to the cryogenic stage in the atom probe instrument, the solution containing NPs is expected to be frozen immediately followed by APT imaging. Based on the acquired mass spectrum (Figure 1b) and reconstructed 3D chemical maps (Figure 1c), the detailed structure and chemical composition of the liquid-NP interface can be analyzed.

Significant findings in this critical domain have been outlined in Figure 1d. It has been revealed for the first time by APT imaging that water ($H_2O$) molecules constitute an electrical double layer (EDL) at water-gold interface, while each layer exhibits distinct mass densities. The thickness of EDL has been proved to be dependent on the charge from the binding between the trisodium-citrate layer and the AuNP. In the bulk water region beyond the AuNP surface, the density of $H_2O$ (ice) appears to be consistent, and no detectable crystalline structures have been found.

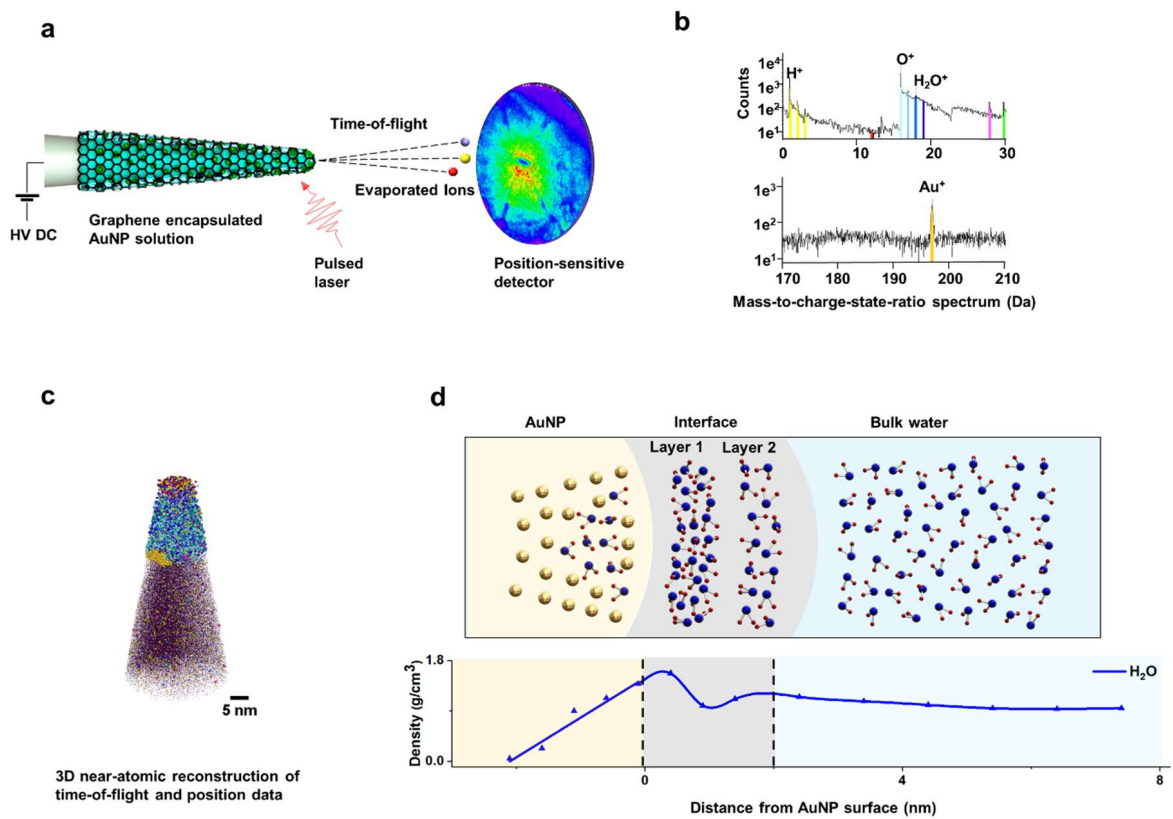

**Figure 1. Overview of the proposed approach for atom probe tomography of liquid-NP interfaces** (a) Schematic of graphene-encapsulated gold nanoparticle (AuNP) specimen probed in a laser-pulsing mode in ultra-high vacuum and cryogenic conditions. (b) An acquired mass-to-charge-state ratio spectrum and (c) the associated 3D chemical map of the field-evaporated volume containing a partial AuNP at near-atomic resolution. (d) Schematic of determined water density as a function of the distance from the AuNP surface at the water-gold interface.

## 2. Results and discussions

### 2.1 Electron microscopy of APT specimens

Prior to APT, SEM and TEM were employed to confirm the AuNP solution encapsulated on the W tip and the final geometry suitable for field evaporation. Notably, graphene encapsulation has been previously shown to mitigate damage due to electron irradiation.[20, 21] Figure 2a shows a SEM image of the graphene-encapsulated AuNP specimen with an end radius in the range of 50 nm. An enlarged region on the specimen tip imaged by TEM (Figure 2b) clearly demonstrates that AuNPs in solution can be encapsulated with graphene. The AuNPs are shown to be adsorbed to the water-tungsten interface, and dispersed due to that a trisodium-citrate layer ($Na_3C_6H_5O_7$) is attached on the NP surface which stabilizes AuNPs in the hydrated state.[22] The average radius of nanoparticles has been determined to be 17.3 nm, which is close to that determined by TEM imaging of the same batch of AuNPs prepared on TEM grid (Figure 2c). The morphology of encapsulated AuNPs suggests that graphene encapsulation process is gentle enough to maintain the integrity of nanoparticles, owing to the ultra-thin characteristic of single graphene layer and the "cushion" effects of the liquid solution involved.[23] Figure 2d demonstrates a series of TEM images at different tilt angles by rotating the sample around a single tilt axis, providing a novel method to prepare liquid specimens for TEM tomography.[24] The twin boundaries of the icosahedral AuNP can be observed in Figure 2d, but the water-gold interface is not clear. This provides further motivations for investigating the liquid-NP interface with APT that is capable of providing detailed chemical maps of a single AuNP in solution at near-atomic resolution.

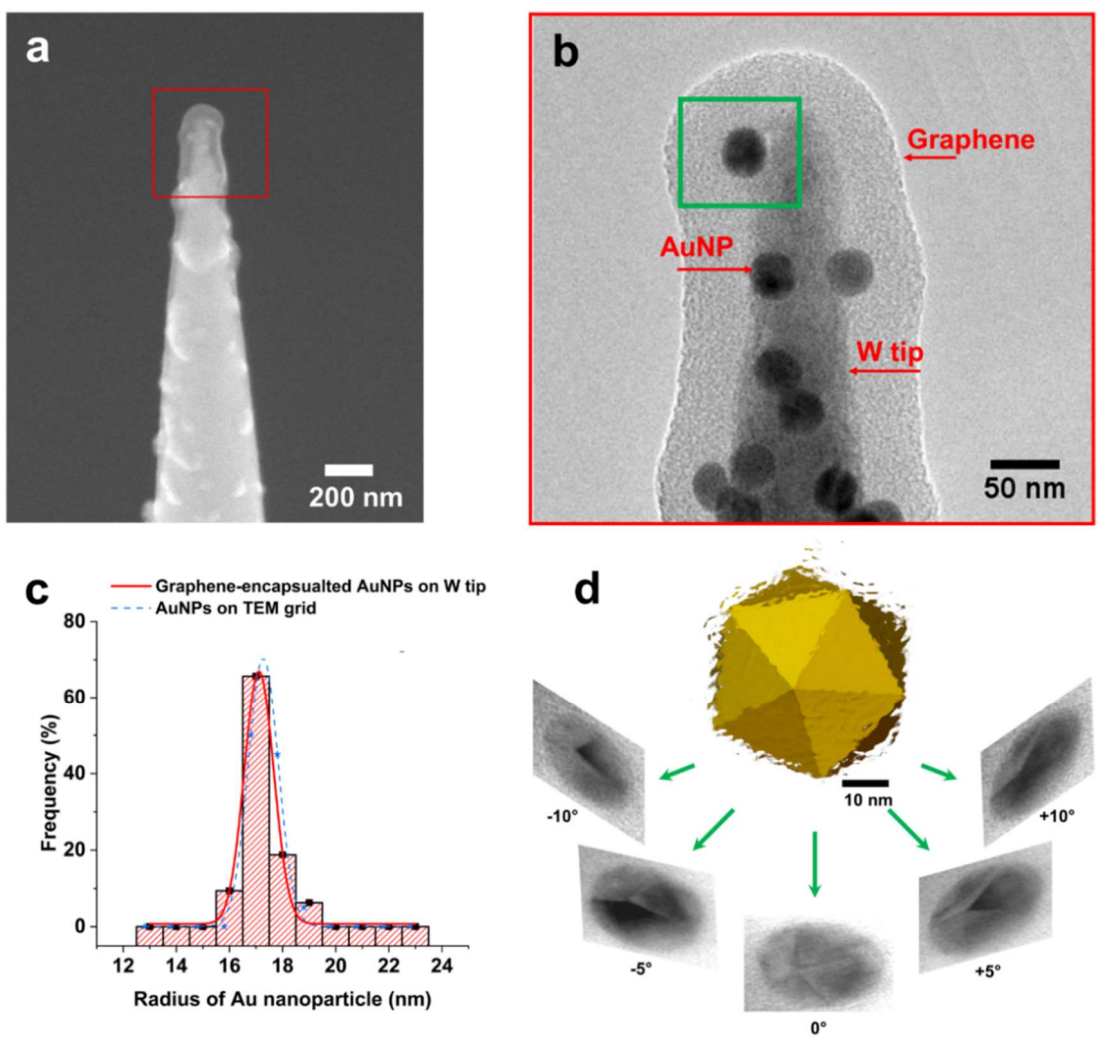

**Figure 2. Electron microscopy of graphene-encapsulated AuNP specimens.** (a) SEM image of graphene-encapsulated AuNP solution on W tip. (b) TEM image of the red box in (a) showing a magnified image of single graphene layer and individual AuNPs in suspension. (c) Particle size analysis comparing TEM imaging of AuNPs on the TEM grid and the graphene-encapsulated AuNPs in solution. (d) TEM images of a single AuNP in solution (green box in (b)) at different tilt angles showing twin boundaries. The schematic of the icosahedral AuNP structure is shown in yellow.

2.2 APT imaging of graphene-encapsulated AuNP specimen

With the final geometry confirmed suitable for field evaporation, the graphene-encapsulated AuNP specimens were then transferred to the atom probe instrument. The specimens were mounted on a cryogenic nano-positioning stage under ultra-high vacuum ($<1\times10^{-11}$ Torr) and cryogenically cooled by a closed-cycle helium cryo-generator. The acquired time-of-flight mass-to-charge-state ratio spectrum (Figure 3a-b) identifies the species of atomic and molecular ions, and the reconstructed 3D chemical maps presented in Figure 3c reveal the spatial positions of these identified ions. Due to that there is no literature regarding the evaporation field difference between Au and $H_2O$ (ice), a comparison of voltage and mass history (Figure S1a) is examined, showing the transition from water section into the AuNP-containing section. It has been found that the voltage increase while probing the section containing AuNP apparently follows the same rate with that while probing the prior water section (Figure S1b). Considering that the AuNP with an area fraction of about 14.9% was exposed to the detector, the evaporation field difference between Au and $H_2O$ is suggested to be small. As such, a compression effect, namely local magnification effect,[25, 26] is expected to be minor for the geometry of reconstructed water. In this study, 3D reconstruction was performed using the tip-profile method that defines the reconstruction radius evaluation based on the TEM image.[27] For example, the geometry of reconstructed solution volume (Figure 3c) is consistent with that from TEM imaging (Figure 2b), suggesting that the local magnification effect to the width and thickness of reconstructed water has been mitigated.[28, 29]

From the acquired mass spectrum, peaks are assigned to atomic and molecular ions containing H, C, O, Na and Au, by referring to the composition of AuNP solution. In this study, peak assignment (Table S1) has been determined based on the ionic species possibly existing within the graphene-encapsulated specimen. An inevitable problem for APT analysis is that different ionic species with the same mass-to-charge-state (Da) values can be assigned to the same peak. In this experiment, the interferences due to mass-to-charge-state overlap, such as $Na^+$ and $NO_2^{2+}$ (23 Da) as well as $COO^+$ and $CO_2^+$ (44Da), are minimized by consulting their spatial locations together with the known composition. For example, if ions at 23 Da and 44 Da are in close proximity of reconstructed AuNP, these ions were identified as $Na^+$ and $COO^+$ coming from the trisodium-citrate layer. Ions with same mass-to-charge-state values but located predominantly at the top of the reconstructed solution volume were determined as $CO_2H_2^{2+}$ and $CO_2^+$ which are possibly originated from the graphene layer.

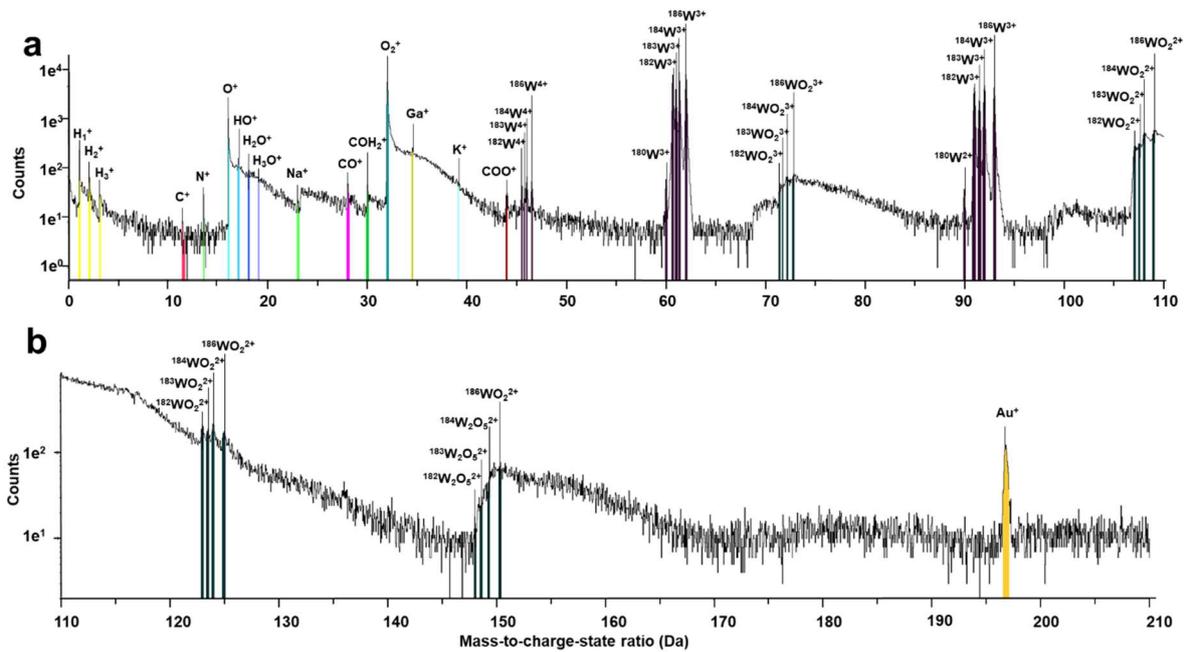

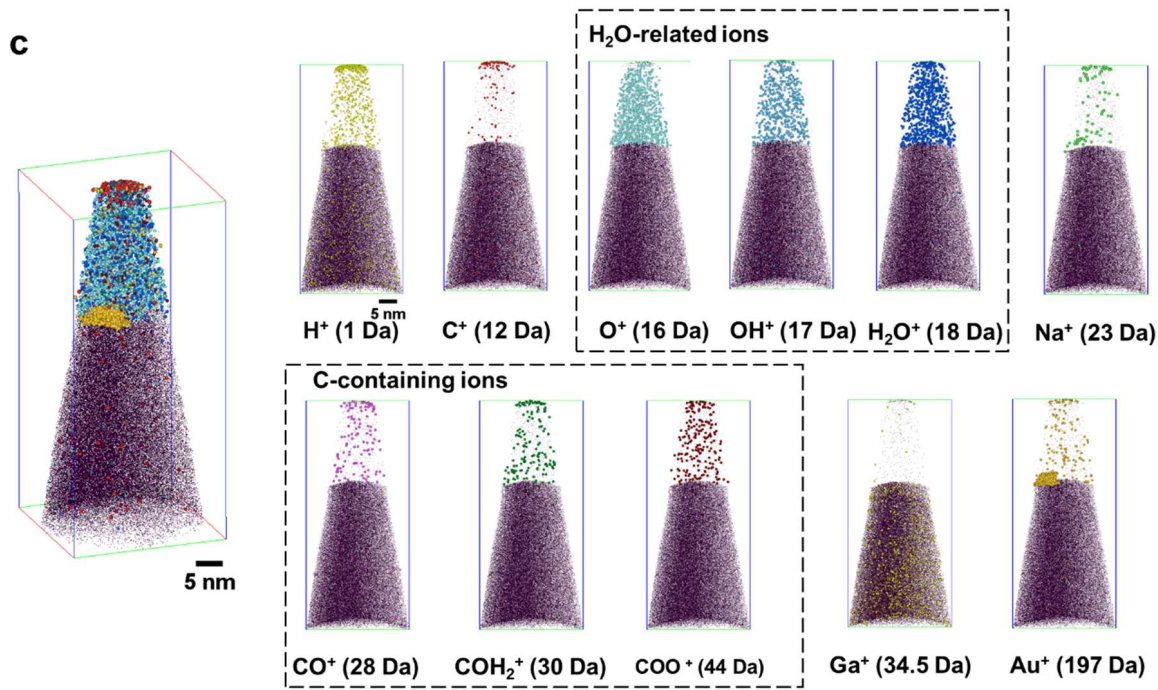

**Figure 3. Acquired mass spectrum and reconstructed 3D chemical maps of graphene-encapsulated AuNP specimen.** The mass spectrum of a graphene-encapsulated AuNP specimen with mass-to-charge-state ratio ranging from (a) 0-110 Da and (b) 110-210 Da. (c) 3D near-atomic reconstruction of the graphene-encapsulated AuNP specimen showing individual maps for different ionic species.

From the reconstructed 3D chemical maps of the APT specimen (Figure 3c), two distinct volumes have been revealed, corresponding to the base W tip and AuNP solution volume on top. Tungsten ions in varied charge states and isotopes ($^{182-186}W^{1-3+}$) are the most abundant in the proportion of detected ions (92%). $Ga^{2+}$ (34.5 Da) ions are only identified inside the W volume since the W tip was pre-sharpened by FIB milling with gallium ions. The W-containing ions were only detected in the region lower than 22 nm from the top of the evaporated volume, suggesting that the ions collected from the top are from the encapsulated AuNP solution. A significant layer of $C^+$ (12 Da) ions has been detected at the top of the specimen which may originate from the graphene membrane. $H_2O$-related ions including $O^+$ (16 Da), $OH^+$ (17 Da), and $H_2O^+$ (18 Da) dominate the solution volume, confirming the encapsulation of AuNP solution with minimal dehydration. The ionization and dissociation of $H_2O$ occurred during field evaporation and ionization. $Na^+$ (23 Da) and C-containing ions including $CO^+$ (28 Da), $COH_2^+$ (30 Da) and $COO^+$ have been detected inside the solution volume, due to that the added trisodium citrate binds with the AuNPs and enhances the stabilization of AuNPs in solution.[22] Highly packed $Au^+$ (197 Da) ions are detected within a confined region, which represents a single AuNP.[14] $H^+$ (1 Da) ions are distributed across both the W and solution volume due to background $H^+$ residuals in the vacuum,[30, 31] similar with $H_2^+$ and $H_3^+$. In the following three repeats of APT imaging with the same batch of AuNP solution, no significant changes of ionic species have been identified in the acquired mass spectra, as shown in Figure S2.

2.3 APT analysis of liquid-NP interface

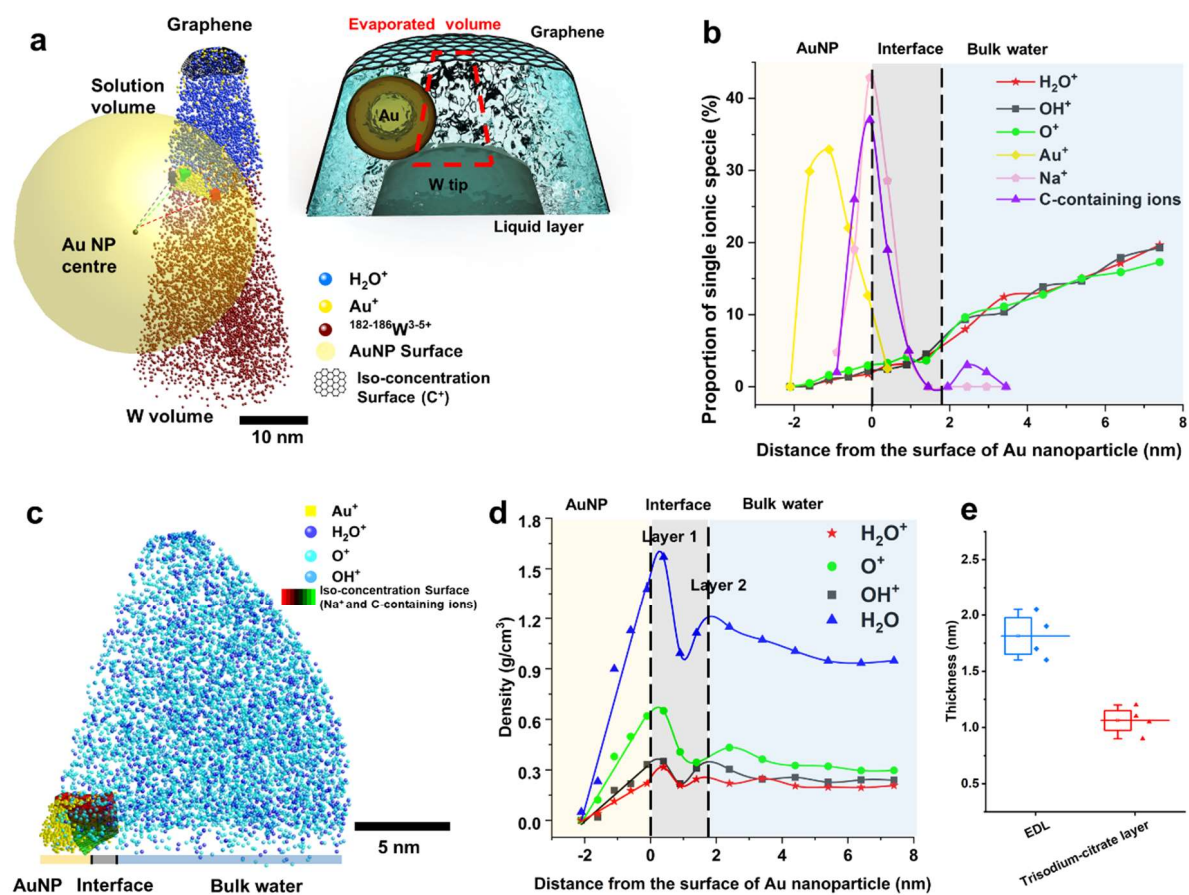

**Figure 4. APT reconstruction of a single AuNP in solution.** (a) Dense local regions of $C^+$ and $Au^+$ ions representing the graphene membrane and AuNP respectively. (b) Proportion of single ionic species across the water-gold interface along the radial direction from the center of the AuNP. (c) Enlarged 3D view of reconstructed water-gold interface, with an iso-concentration surface fitted to the $Na^+$ and C-containing ions to locate the trisodium-citrate layer (20 at.% $Na^+$ and C-containing ions). (d) Determined density of $H_2O$ and $H_2O$-related ions as a function of the distance from the estimated surface of AuNP. (e) Error bars represent

the range of determined thickness of electrical double layer (EDL) and trisodium-citrate layer in four experimental runs (n=4).

By density-based cluster identification,[32] two species of ions, such as $C^+$ (12 Da) and $Au^+$ (197 Da), have been found to be highly concentrated in local regions of solution volume. 84% of $C^+$ ions are distributed in the top layer of the solution volume with a layer thickness of approximately 0.5 nm (Figure 4a). This layer is considered to represent the single graphene layer. The majority of acquired ions other than $C^+$ are located beneath the identified graphene layer, revealing the significant sealing capability of graphene membrane.[33] As demonstrated in Figure 4a, the region of highly concentrated $Au^+$ ions represents a significant portion of a single AuNP, while the surface as well as the center of this nanoparticle can be estimated through fitting a spherical surface with the minimum square error from the $Au^+$ ions at the water-gold interface.[34, 35] The distance from the center to the surface is determined to be 17.6 nm, consistent with the average radius from TEM imaging (Figure 2c).

Based on the hypothesized spherical coordinate with an origin of reconstructed AuNP surface, the number of individual ionic species can be assessed for each 0.4 nm interval along the radial axis from the center (Figure S1c), as well as the proportion of single ionic specie that was calculated from the number of an individual ionic specie within a given shell region over the total number of the same ionic specie detected (Figure 4b). 94% of $Au^+$ ions are distributed inside the AuNP surface. Well above the AuNP surface, 80% of $H_2O$-related ions ($O^+$, $OH^+$, and $H_2O^+$) have been detected beyond a distance of 1.8 nm, and this region is considered as bulk water. Between the AuNP and bulk water lies the water-gold interface, where an abundant species of ions including $Au^+$, $H_2O$-related ions, $Na^+$ and C-containing ions (Figure 4b) are present. Approximately 4% of $Au^+$ ions are detected at the interface, as well as the number of $H_2O$-related ions has been found to fluctuate (Figure S1c). Since the $Na^+$ and C-containing ions ($CO^+$, $COH_2^+$ and $COO^+$) are both considered to originate from the trisodium citrate, a surface based on the concentration of these ions can be constructed to represent the trisodium-citrate layer (Figure 4c). This surface is shown to cover the entire reconstructed water-gold interface, verifying at a near-atomic resolution that the trisodium citrate is uniformly distributed on the AuNP surface with the major connecting bonds of $Au^+$-$COO^+$/$COH_2^+$ and $Au^+$-$Na^+$. As the amount of C-containing ions is larger than that of $Na^+$ (Figure S1c and S1e), the AuNP is postulated to be negatively charged. This result agrees with the previously reported adlayer structure obtained by scanning probe microscopy, AFM and solid-state nuclear magnetic resonance (NMR).[36-38] The experimentally determined thickness of the trisodium-citrate layer in this study is approximately 1 nm (n=4) (Figure 4e) based on the APT reconstruction, which is comparable to the estimated range of 0.8-1 nm based on the theoretical length of dithiols.[22] $H_2O$-related ions have been located on both sides of this adlayer, demonstrating the atomic-resolution view of trisodium-citrate layer in the solution.

Figure 4d reveals the density of $H_2O$ based on the mass density of individual $H_2O$-related ions, including $O^+$, $OH^+$, $H_2O^+$, from the NP interior to regions across the liquid-NP interface. One notable finding is that $H_2O$ molecules have formed an EDL at the interface. As revealed in Figure 4d, the mass density of $H_2O$ in the first layer is significantly higher than the density in the second layer. The formation of the EDL results from that the presence of trisodium-citrate layer charges the AuNP.[22, 39] The overall thickness of EDL has been found to be approximately 1.8 nm (n=4) (Figure 4e), which is larger than the reported thickness (~ 1 nm) of interfacial water near gold electrodes at a positive bias measured by x-ray absorption spectroscopy.[16] This is due to that the intensity of the electric field generated by the positive bias, namely the voltage, is larger than the one generated by the binding between the trisodium-citrate layer and AuNP, and a more intensive electric field can significantly enhance the absorption of $H_2O$ molecules.[4] Considering that the electric intensity and the electric potential both can affect the thickness of the EDL,[4, 16] the controllable thickness of EDL at the liquid-NP interface can be achieved by tuning the ligand on the NP surface in order for varied charge states. Within the bulk water region, the mass density of $H_2O$ appears to be consistent, and the determined density of 0.95 g/cm$^3$ is slightly larger than the reported value of 0.934 g/cm$^3$ at 93 K.[40] Furthermore, the density change of $H_2O$ can possibly be pressure induced due to graphene encapsulation.[41]

Another finding is that $H_2O$-related ions have been found inside the AuNP (Figure 4b and 4d). A couple of possible causes exist. Firstly, the lattice plane interspacing of Au (~2.4 Å) is comparable to the theoretical diameter of $H_2O$ molecules,[42] and existing defects on the AuNPs surface may have enlarged the spacing, which provides opportunities for the existence of $H_2O$ interior AuNPs. Secondly, although the evaporation field difference between Au and water ice has been shown to be minor based on the voltage and mass history (Figure S1a), trajectory aberration and trajectory overlap due to local magnification effect still can play a role.[25, 26] In particular, the trajectory aberration is typically limited to be 0.3 nm laterally,[9] and the resultant interfacial overlapping is limited to be approximately 0.5 nm if the evaporation field difference is less than 20%.[43, 44]

The results of this study have suggested that the thickness of EDL is dependent on the charge from binding between NP and the ligand, which provides an alternative route for developing double-layer capacitors with controllable capacitance.[45-48] In addition, the proposed method of APT imaging of liquid-NP interface will support the exploration of functionalized NPs and their interactions in different solutions. The capability of near-atomic resolution will also help to expand the knowledge of the diverse crystalline and amorphous phases of water ice,[49, 50] inside the miniaturized "pressure chamber" constructed by graphene encapsulation.

## 3. Summary

A novel approach is reported to explore the structure and chemical composition at the liquid-NP interface with near-atomic resolution, which overcomes the longstanding challenges in high-resolution chemical mapping of this critical domain. By introducing graphene encapsulation, the solution containing NPs has been encapsulated on pre-sharpened metal tips, and the final geometry remains suitable for APT imaging. In this study, a large amount of $H_2O$-related ions including $O^+$, $OH^+$ and $H_2O^+$ have been reconstructed, together with the AuNP represented by a highly packed region of $Au^+$. The trisodium-citrate layer on the AuNP surface has also been observed based on the $Na^+$ and C-containing ions ($CO^+$, $COH_2^+$ and $COO^+$). It is the first demonstration that $H_2O$ molecules form an EDL with an overall thickness of approximately 1.8 nm at the water-gold interface. The formation of EDL is due to that the binding of trisodium-citrate layer charges the AuNP. In the bulk water region, $H_2O$ molecules appear to be uniformly packed without detectable crystalline structures. In summary, a complete 3D chemical map of the water-gold interface at near-atomic resolution has been reconstructed, starting from the interior of AuNP to the water-gold interface and finally ending in the bulk water region. The proposed technology is expected to be a unique approach in the future to study materials in a controlled environment, such as the interface between the electrode and liquid electrolyte, as well as biological processes driven by macromolecules and drug solutions.

## Acknowledgments

This study was partly funded by the Australian Research Council (DP180103955). This work was performed in part at the Melbourne Centre for Nanofabrication (MCN), Victorian Node of the Australian National Fabrication Facility (ANFF). Also, the authors acknowledge the use of facilities within the Monash Centre for Electron Microscopy (MCEM), Monash Ramaciotti Cryo-EM platform, Monash Campus Cluster (MCC), Deakin University's Advanced Characterization Facility, and Nanjing University of Science and Technology's Chinese Centre of Excellence for Atom Probe Tomography.